\documentclass[epj]{svjour}
\usepackage{graphics}
\begin{document}
\title{Nuclear mass form factors from coherent photoproduction of \\$\pi^0$ mesons  
}
\author{B. Krusche\inst{1}
}                     
\offprints{Bernd.Krusche@unibas.ch}          %
\institute{Department of Physics and Astronomy, University of Basel,
           Ch-4056 Basel, Switzerland 
           }
\date{Received: date / Revised version: date}
%
\authorrunning{B. Krusche}
\titlerunning{Nuclear mass form factors}

\abstract{Data for coherent photoproduction of $\pi^0$ mesons from nuclei
($^{12}$C, $^{40}$Ca, $^{93}$Nb, $^{nat}$Pb), recently measured with the TAPS 
detector at the Mainz MAMI accelerator, have been analyzed in view of the mass 
form factors of the nuclei. The form factors have been extracted in plane wave 
approximation of the $A(\gamma ,\pi^0)A$ reaction and corrected for final
state interaction effects with the help of distorted wave impulse approximations.
Nuclear mass rms-radii have been calculated from the slope of the form factors 
for $q^2\rightarrow 0$.
Furthermore, the Helm model (hard sphere form factor folded with Gaussian)
was used to extract diffraction radii from the zeroes of the form factor
and skin thicknesses from the position and height of its first maximum.
The diffraction radii from the Helm model agree with the corresponding
charge radii obtained from electron scattering experiments within their 
uncertainties of a few per cent. The rms-radii from the slope of the form 
factors are systematically lower by up to 5\% for PWIA and up to 10\% for DWIA.
Also the skin thicknesses extracted from the Helm model are systematically
smaller than their charge counter parts.
} 
\maketitle
\section{Introduction}
\label{sec:1}

Charge and matter densities are among the most fundamental properties of
atomic nuclei. Nuclear charge distributions have been intensively studied with 
elastic electron scattering and via the spectroscopy of X-rays from muonic
atoms (see e.g. \cite{deJager_74,deVries_87,Frois_91,Fricke_95}). 
These experiments profit from the full understanding of the electromagnetic
interaction. Analyses of the distributions in the frameworks of different 
models have extracted characteristic parameters like charge radii, 
skin thicknesses, or the central depression of the charge density
with high precision  \cite{Friedrich_82,Friedrich_86}. However, all these 
properties are only related to the distribution of the protons in the nucleus.
The electromagnetic interaction provides only very limited 
information on the neutron. Therefore, the extraction of neutron distributions,
respectively nuclear matter distributions  (i.e. the sum of proton and neutron 
density) is much less straight forward. Some results for specific single 
neutron orbits have been obtained with elastic magnetic electron scattering, 
making use of the magnetic form factor of the neutron 
\cite{Sick_77,Platchkov_82}. However, most experimental methods use hadron
induced reactions such as for example $\alpha$, proton, pion or kaon 
scattering from nuclei. The analysis of such reactions requires scattering
theories for strongly interacting particles, with all their uncertainties.    
An overview over the different methods can be found in \cite{Batty_89}.
The situation is such, that a systematic analysis of the nuclear matter
distributions is still missing and in many cases the spread between the 
results obtained with different probes is still larger than the predicted
differences in the proton and neutron distributions, which are on the order of
0.05 fm - 0.2 fm for the rms radii of heavy nuclei \cite{Pomorski_97}. 

The present paper summarizes the analysis of recent experimental results for 
the coherent photoproduction of $\pi^0$ mesons \cite{Krusche_02} in view of
nuclear matter distributions. This reaction is particularly 
attractive as a complementary method for the study of nuclear matter 
distributions of stable nuclei. As discussed below, in the energy region of
interest, protons and neutrons contribute identically with the same amplitude.
Furthermore, in contrast to hadron induced reactions it is not restricted to 
the nuclear surface but probes the entire nuclear volume. In this sense it is
the ideal reaction to test the matter distribution in the bulk of a nucleus.
A first attempt to determine nuclear mass radii
with this method was made by Schrack, Leiss and Penner in 1962 
\cite{Schrack_62}. However, at that time the achievable experimental precision 
was very limited and much inferior to the results from hadron 
induced reactions which profit from the large cross sections characteristic for
the strong interaction. Subsequently, an attempt was made to measure nuclear 
matter radii via the coherent photoproduction of $\rho^o$ mesons 
\cite{Alvesleben_70}. However, due to the experimental difficulties and the
previously not well developed theoretical tools for the extraction of the form 
factors from coherent photoproduction reactions, the method was never
systematically explored. On the experimental side the progress made in
accelerator and detector technology during the last fifteen years has
considerably enhanced our possibilities for the study of photon induced
meson production reactions. In particular, the new generation of quasi 
continuous beam electron accelerators like CEBAF in Newport News, ELSA in Bonn, 
ESRF in Grenoble, MAMI in Mainz and SPring8 in Osaka are all equipped with
state-of-the-art tagged photon facilities and highly efficient 
detector systems, most of them with almost $4\pi$ solid angle coverage.
Profiting from this developments, recently much more precise cross section
measurements for coherent $\pi^0$ photoproduction from carbon, calcium, niobium
and lead nuclei became available \cite{Krusche_02}. At the same time, 
progress was also made in the theoretical understanding of these reactions.
Modern calculations treating the elementary process in the framework of unitary
isobar models, incorporating final state interaction (FSI) in distorted wave
impulse approximation, and including in-medium effects of the $\Delta$-isobar    
with phenomenological self-energies have become available \cite{Drechsel_99}.
The purpose of this paper is thus to explore the currently achievable accuracy
in the determination of the nuclear mass distributions from coherent $\pi^0$
photoproduction. This is also done in view of the possibility of further 
improvements in the data quality. Although the data from \cite{Krusche_02},
which are the basis of the present analysis, are the most precise 
results for coherent $\pi^0$-photoproduction from heavy nuclei available 
so far, they have been measured with an early stage 
of the TAPS detector \cite{Novotny_91,Gabler_94} covering only $\approx 20$\% 
of $4\pi$. This resulted in typical detection efficiencies for $\pi^0$ mesons 
on the order of only a few per cent and an imperfect suppression
of incoherent contributions from excited nuclear states. With the availability 
of $4\pi$ detector systems like the combined Crystal Ball/TAPS setup, operating
now at the MAMI accelerator, more precise data will become available in the 
future.

In this paper, we will limit the discussion of the mass distributions to 
the extraction of rms radii from the slope of the form factors and to the 
extraction of diffraction radii and surface extensions in the Helm model.
However, the analysis of the full matter distributions in the same model 
independent way as for the charge distributions should become feasible
for the next generation of experiments.   
   
\section{Coherent $\pi^0$ photoproduction from nuclei}
\label{sec:2}
Coherent photoproduction of $\pi^0$-mesons from a nucleus with mass number
$A$ is the reaction
\begin{equation}
\gamma+A(gs)\rightarrow A(gs) + \pi^0
\label{eq:1}
\end{equation}
where $A(gs)$ is a nucleus in its ground state. It can be experimentally
separated from breakup processes, where nucleons are removed from the nucleus,
via its characteristic two-body kinematics \cite{Krusche_02}. The theoretical 
treatment is much more straight forward than for incoherent pion production 
reactions, since for initial and final state only ground state properties of the
nucleus are needed. 

In general, the isospin structure of the elementary process of $\pi^0$ 
photoproduction from the nucleon is given by
\begin{eqnarray}
\label{eq:2}
A(\gamma p\rightarrow\pi^o p) & = &
+\sqrt{\frac{2}{3}}\;A^{V3}+\sqrt{\frac{1}{3}}(A^{IV}-A^{IS})\\
A(\gamma n\rightarrow\pi^o n) & = &
+\sqrt{\frac{2}{3}}\;A^{V3}+\sqrt{\frac{1}{3}}(A^{IV}+A^{IS})\;,\nonumber
\end{eqnarray}
where $A^{IS}$, $A^{IV}$, and $A^{V3}$ are the isoscalar, isovector, and
total isospin changing parts of the total amplitude. However, at incident 
photon energies in the range of interest in this work (200 - 350 MeV) the
reaction is completely dominated by the photo excitation of the $\Delta(1232)$ 
resonance (see e.g. \cite{Krusche_03}). Since this is an isospin $I=3/2$ state
only the isospin changing vector component $A^{V3}$ can contribute, so that for 
the $\Delta$ excitation
\begin{equation}
A(\gamma p\rightarrow\pi^o p)
=
A(\gamma n\rightarrow\pi^o n)\;\;.
\label{eq:3}
\end{equation}
Detailed investigations of coherent and breakup photoproduction of $\pi^0$
mesons from the deuteron \cite{Krusche_99,Darwish_03,Krusche_03} have 
confirmed that the elementary cross sections for protons and neutrons
are equal.
This means that apart from small background contributions (nucleon Born terms)
protons and neutrons contribute with the same amplitude to coherent $\pi^0$
photoproduction from nuclei, so that this reaction is indeed sensitive to the 
distribution of nucleons rather than to the distribution of charge in the
nucleus.

In the most simple plane wave impulse approximation (PWIA)
the coherent cross section from spin zero nuclei can be written as
\cite{Drechsel_99,Krusche_02}
\begin{equation}
\frac{d\sigma_{PWIA}}{d\Omega}(E_{\gamma},\Theta_{\pi})
=
\frac{s}{m_N^2}A^2
\frac{d\sigma_{NS}}{d\Omega^{\star}}(E^{\star}_{\gamma},\Theta_{\pi}^{\star})F^2(q)
\label{eq:4}
\end{equation}  

\begin{equation}
\frac{d\sigma_{NS}}{d\Omega^{\star}}(E^{\star}_{\gamma},\Theta_{\pi}^{\star})
=
\frac{1}{2}\frac{q_{\pi}^{\star}}{k^{\star}}
|{\cal{F}}_2(E_{\gamma}^{\star},\Theta_{\pi}^{\star})|^2 
sin^2(\Theta_{\pi}^{\star}),
\label{eq:5}
\end{equation}
where $E_{\gamma}$ and $\Theta_{\pi}$ are incident photon energy and pion 
polar angle in the {\em photon-nucleus} cm-system, $m_N$ is the nucleon mass,
$q(E_{\gamma},\theta_{\pi})$ the momentum transfer to the nucleus, 
and $F(q)$ the nuclear mass form factor. The total energy $\sqrt{s}$ of the 
photon-nucleon pair, the photon energy and momentum 
$E_{\gamma}^{\star}$, $k^{\star}$, and the
pion angle and momentum $\Theta_{\pi}^{\star}$, $q_{\pi}^{\star}$ in the 
{\em photon-nucleon} cm-system can be evaluated from the average momentum 
$\vec{p_N}$ of the nucleon in the factorization approximation 
$\vec{p}_N=\vec{q}(A-1)/2A$.
The spin-independent elementary cross section $d\sigma_{NS}/d\Omega$
is calculated from the isospin average (for $I\neq 0$ nuclei weighted
with N,Z) of the standard Chew-Goldberger-Low-Nambu (CGLN) amplitude 
${\cal{F}}_2$ \cite{Chew_57} taken from \cite{Drechsel_99a}.
The extraction of the form factor $F(q)$ from the differential cross sections
in this approximation is straight forward and used below for a first
approximative determination of the mass radii. 

\begin{figure}[h]
\centerline{\resizebox{0.5\textwidth}{!}{%
  \includegraphics{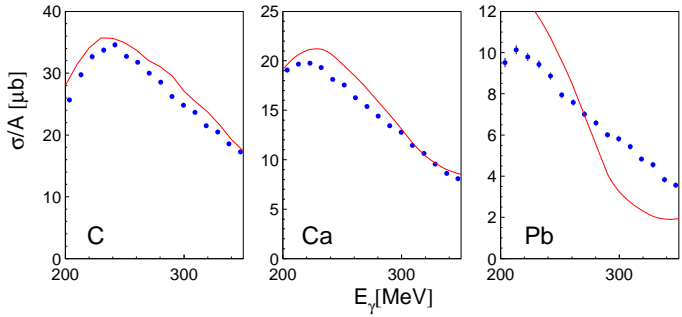}
}}
\caption{Total cross section for coherent $\pi^o$ photoproduction from
carbon, calcium and lead \cite{Krusche_02} compared to the model results from
\cite{Drechsel_99}.}
\label{fig:total}       
\end{figure}

It is well known, that final state interaction effects can have a
significant impact on the pion production cross sections. The available model 
calculations have been compared to the measured differential cross
sections in \cite{Krusche_02}.
The distorted wave impulse approximation in momentum space with additional 
$\Delta$ in-medium effects by Drechsel et al. \cite{Drechsel_99} gave the best 
agreement with the data. The results from this
model are therefore used in the present work for the correction of FSI effects
in the extraction of the nuclear form factor. It should be emphasized that the 
model \cite{Drechsel_99} was not adjusted to the nuclear data under discussion.
The free parameters for the $\Delta$-nucleus phenomenological self energy  
were fitted to coherent $\pi^0$ photoproduction from $^4$He and not modified for
the heavy nuclei. The typical agreement between data and model results for the 
total cross sections is shown in fig. \ref{fig:total}. It is quite good for 
carbon and calcium, but less so for lead. However, also for lead important 
features for this analysis, like the position of the diffraction minima, are 
very well reproduced (see \cite{Krusche_02} for a detailed discussion). 
For the present analysis of form factors only the relative shifts of the
position of the minima and cross section ratios between the PWIA calculation 
and the full model from \cite{Drechsel_99} are used. Such calculations are
presently not available for the nucleus $^{93}$Nb. However, the corrections
for the shift of the minima and the slope at small $q$ can be approximated
from lead by normalizing the position of the first diffraction minimum.
Strictly speaking, also the above PWIA approximation for spin zero
nuclei is not valid for this $J=9/2$ odd-even nucleus. However, since it was 
shown in \cite{Krusche_02} that the measured cross sections scale in the same 
way as for the spin zero nuclei (the contribution from the odd nucleon
is not significant) we have kept it in the analysis. The systematic uncertainty
is of course larger than for the other nuclei.

\begin{figure}[th]
\centerline{\resizebox{0.5\textwidth}{!}{%
  \includegraphics{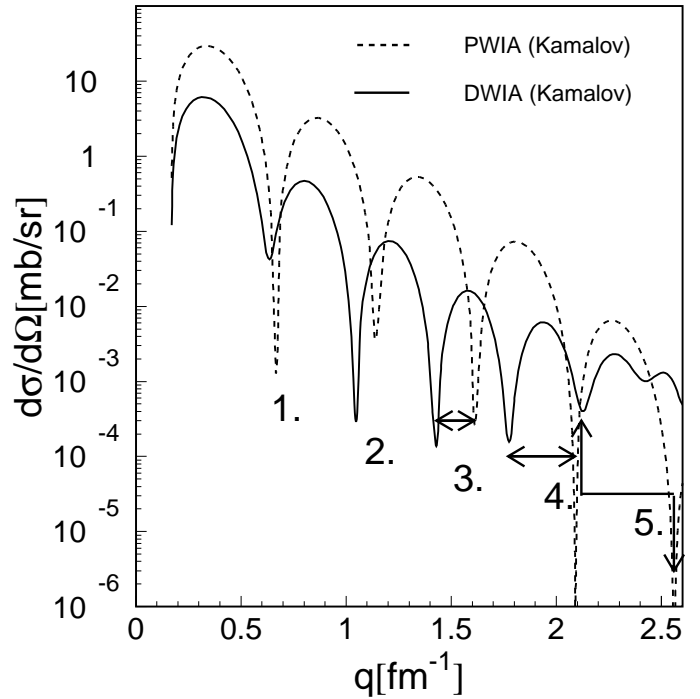}
}}
\caption{PWIA and DWIA calculations for coherent $\pi^0$ photoproduction from
lead for an incident photon energy of $E_{\gamma}$=290 MeV
\cite{Drechsel_99,Krusche_02}. The shift of the 
diffraction minima due to FSI is indicated.}
\label{fig:sabit}       
\end{figure}

An example for the influence of FSI on the angular distribution from lead
\footnote{Note: differential cross sections $d\sigma /d\Omega$ are given 
as function of $q$ throughout this paper, which for fixed $E_{\gamma}$
is a unique function of $\theta_{\pi}$. This has the advantage that the 
position of the diffraction minima is approximately independent on 
$E_{\gamma}$.}
for an incident photon energy of 290 MeV
is shown in fig. \ref{fig:sabit} where the PWIA and DWIA calculations for the
differential cross section are plotted versus the momentum transfer to the
nucleus.
The main effects at these incident photon energies are a reduction
of the plane wave cross section and a shift of the position of the diffraction 
minima due to the pion - nucleus potential. The latter is more pronounced 
for the higher order maxima.

For the interpretation of the form factor results it is of interest if
final state interaction effects, in particular pion absorption, are so strong
that like in hadron induced reactions effectively only the nuclear surface is
tested.
\begin{figure}[h]
\centerline{\resizebox{0.5\textwidth}{!}{%
  \includegraphics{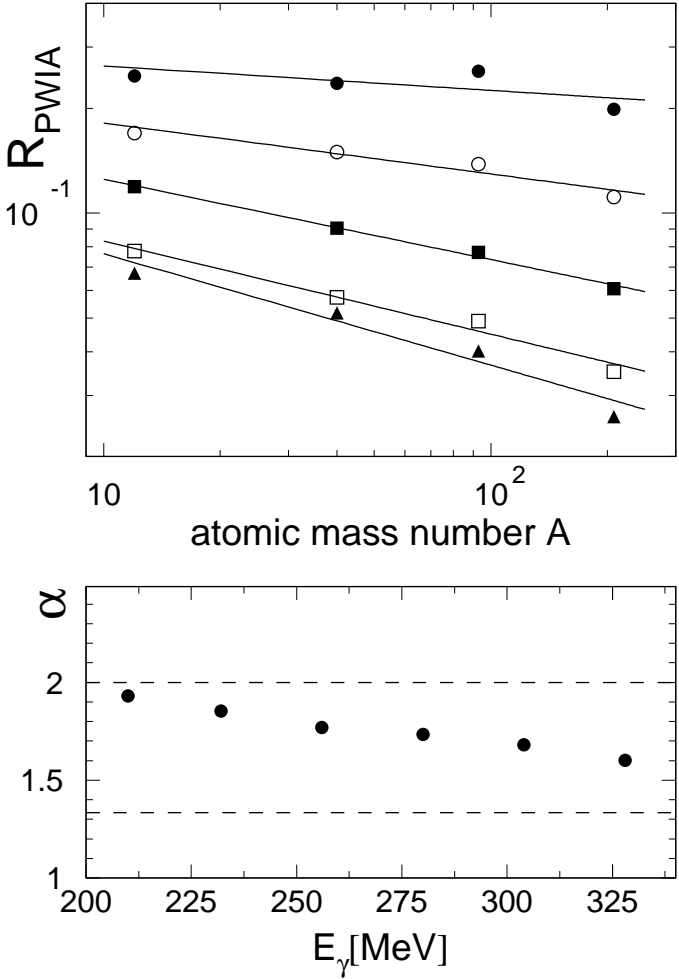}
}}
\caption{Upper part: $A$-dependence of $R_{PWIA}$ at $q=0.5q_1$ for incident
photon energies of 210, 230, 255, 280, and 305 MeV (from top to bottom).
Lower part: fitted coefficients $\alpha$ of the mass dependence.}
\label{fig:adep}       
\end{figure}

An indication for the strength of these effects can be obtained from the scaling
of the cross sections with mass number in a similar way as in
\cite{Krusche_04a,Krusche_04b} for incoherent meson production ($\pi$, $\eta$)
from nuclei. The production cross sections without FSI for incoherent processes
scale with the number of nucleons $A$. The measured cross sections scale like 
$A^{2/3}$, which implies strong FSI corresponding to a small mean free path of 
the mesons so that effectively only the nuclear surface contributes.
The cross section for coherent $\pi^0$ production in plane wave without
FSI scales like $A^2$, since the amplitude is proportional to $A$.
If only the surface would contribute one would expect a scaling with
$A^{4/3}$. The observed scaling for the {\em total} coherent cross section 
does not even reach $A^{2/3}$ \cite{Krusche_05}. However, one must take into
account the influence of the $sin^2(\Theta)F^2(q)$ term in the PWIA cross 
section, which contributes to the $A$-dependence of the total cross section
(see eq. (\ref{eq:5})). The nuclear form factor at $q$ normalized to the
position of the first diffraction minimum, $q=q_1(A)$, is almost  
independent on $A$ in the region before the first diffraction minimum
(see fig. \ref{fig:pwia}. Consequently, the mass number scaling 
of the coherent cross sections can be obtained by fitting the ratio $R_{PWIA}$ 
(see eq. (\ref{eq:6})) at a constant value of $q/q_1<1$ with the ansatz
\begin{equation}
R_{PWIA}(A)\propto A^{\alpha -2}
\end{equation}
The result for $q=q_1(A)/2$ (i.e. approximately in the 0-th maximum 
\footnote{Note: the maxima of the differential cross section are labeled
0,1,2,... ,the maxima of the form factor 1,2,3,... . In both cases maximum
1 follows the first diffraction minimum.} of the differential cross section) 
is shown in fig. \ref{fig:adep}.
At the lowest investigated incident photon energies around 200 MeV the scaling
is very close to $A^2$, indicating almost negligible pion absorption.
At higher incident photon energies FSI effects become more important. 

\section{Extraction of mass radii}
\label{sec:3}

\subsection{rms-radii}
\label{ssec:11}

\begin{figure}[th]
\centerline{\resizebox{0.5\textwidth}{!}{%
  \includegraphics{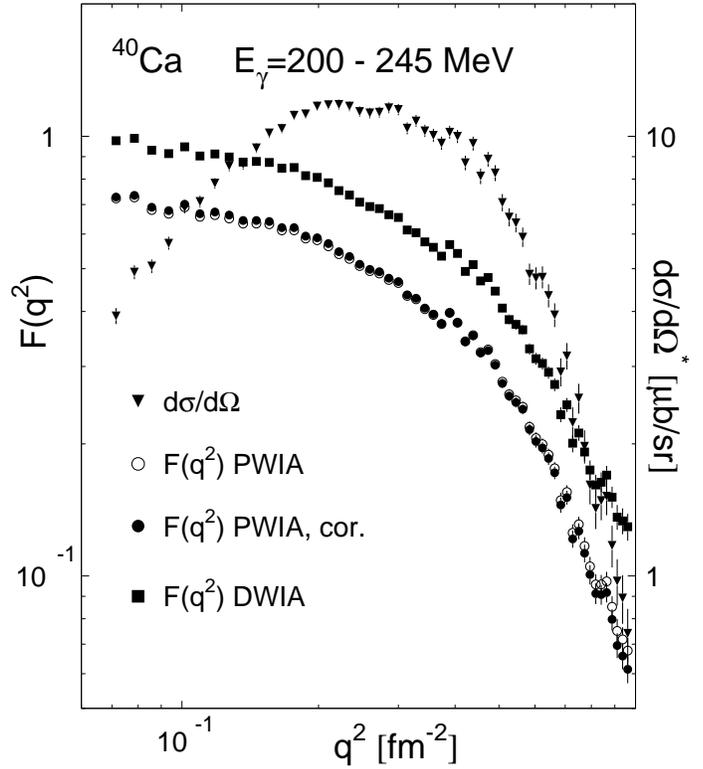}}}
\caption{Extraction of 
the nuclear form factor for $^{40}$Ca. Triangles: 
measured differential cross section (axis labeling at right hand side).
Open circles: PWIA approximation of form factor from 
eqs. (\ref{eq:4}),(\ref{eq:5}).
Filled circles: PWIA approximation after correction of angular resolution
effects. Filled squares: after correction of FSI effects (DWIA approximation). 
}
\label{fig:ana}       
\end{figure}

The determination of the root-mean-square (rms) radii requires the extraction of 
the nuclear form factor $F(q^2)$ from the angular distributions. This is done
in three steps as demonstrated in fig. \ref{fig:ana} for $^{40}$Ca for one range
of incident photon energy. 
In the first step the form factor is extracted from the data in plane wave
approximation, i.e. all FSI effects are neglected.
This is done in the same way as in \cite{Krusche_02}
\footnote{Note: eq. \ref{eq:6} is applied to the data in the finest possible
binning of incident photon energies, defined by the resolution of the tagging
spectrometer (roughly 2 MeV). The corresponding form factor results
are then averaged over larger energy regions for statistical reasons.} 
using 
\begin{equation}
\left.F^2(q)\right|_{PWIA}= R_{PWIA}=
(\frac{d\sigma_{exp}}{d\Omega})/
[\frac{s}{m_N^2}A^2
(\frac{d\sigma_{NS}}{d\Omega^{\star}})]
\label{eq:6}
\end{equation} 
with $d\sigma_{NS}/d\Omega^{\star}$ from eq. (\ref{eq:5}).

In the next step a correction is applied for the finite angular resolution of 
the experiment. It was determined with Monte Carlo simulations that the detector
response for $\pi^0$ mesons in the kinematical regime of interest corresponds
to a Gaussian with a FWHM of 4$^o$ for the pion cm angles. The DWIA model
calculation was folded with this response function and the data were corrected
with the ratio of folded and unfolded calculation. This correction is only
significant for niobium and lead.

In the final step the ratio of PWIA and DWIA cross sections obtained from the
model calculations is used to correct the form factor for FSI effects. The $q$
dependence (not the absolute values) of the PWIA and DWIA calculations is
in most cases similar for small $q$, nevertheless this correction introduces
an additional model dependence into the analysis. We have therefore extracted
the mass radii from the form factors with and without this correction in order
to get an estimate for the systematic uncertainties. 

Once the form factor has been determined, the rms-radius can be extracted 
without further model assumptions from the slope of the form factor for 
$q^2\rightarrow 0$ via
\begin{equation}
F(q^2)=1-\frac{q^2}{6}r_{rms}^2+{\cal{O}}(q^4),
\label{eq:7}
\end{equation}
which is done in the usual way by fitting a polynomial
\begin{equation}
F(q^2)=\sum_{n=0}^{N}(-1)^{n}a_{n}q^{2n}
\label{eq:8}
\end{equation}
to the data. The rms-radius is then given by 
\begin{equation}
r_{rms}=\sqrt{6a_1/a_0},
\label{eq:9}
\end{equation}
where for a correctly determined form factor $a_0$ should be unity.
This is of course not true for the form factors extracted in PWIA
approximation without correction for FSI effects. However, also
the form factors extracted in DWIA approximation differ in most
cases somewhat from $F(q^2)\rightarrow 1$ for $q^2\rightarrow 0$
(see figs. \ref{fig:rms_1},\ref{fig:rms_2}).
This can be due to systematic uncertainties in the absolute normalization
of the measured cross sections (up to 10\% overall, up to 3\%
relative between different nuclei, see ref. \cite{Krusche_02}) or due to
an imperfect correction of the FSI effects (systematic uncertainties
in the models). Therefore $a_0$ is kept free in all fits to account
for these effects. 

A final remark must be made to the comparison of the mass form factors, 
extracted in this way, to the nuclear charge form factors. In the latter case
the charge distribution of the nuclei is tested. Due to the charge form factor
of the proton the distribution of (point-like) protons $F_{pc}$ in the nucleus
is given by
\begin{equation}
F_{pc}(q)=\frac{F^{ch}(q)}{F_p^{ch}(q)}
\label{eq:12}
\end{equation}
where $F^{ch}$ is the nuclear charge form factor and $F_p^{ch}$ is the charge
form factor of the proton for which we take the dipole form factor
\begin{equation}
F_p^{ch}=\left(1+\frac{q^2}{18.234 \mbox{fm}^{-2}}\right)^{-2}\;\;.
\label{eq:dipole}
\end{equation}
On the other hand, coherent pion production, which proceeds
through the excitation of the nucleon to the $\Delta$ resonance, is testing 
the distribution of point-like nucleons in the nucleus. Therefore for the 
comparison of the mass and charge radii the rms charge radius 
$r_{rms}^p$ = 0.862 fm of 
the proton was subtracted in quadrature from the nuclear charge rms-radii
$r_{rms}^{ch}$
to give the rms-radius $r_{rms}^{pc}$ for the distribution of point-like
protons
\begin{equation}
r_{rms}^{pc}=\sqrt{(r_{rms}^{ch})^2-(r_{rms}^p)^2}\;\;\;.
\label{eq:13}
\end{equation}
For the comparison of skin thicknesses the charge form factors were divided 
by the proton charge form factor.

\subsection{Form factors in the Helm model}
\label{ssec:21}
The extraction of the rms radii from the slope of the form factor has the
advantage that no model of the form factor itself is needed. However, models 
of the form factor which relate for example the radii to the position of the 
diffraction minima allow a much better control of systematic effects like the 
DWIA corrections. Furthermore, additional information can be gained from
such models. The rms radius alone has for example no information about the
extension of the surface zone of the nuclei. A good example are the charge
rms radii of $^{40}$Ca and $^{48}$Ca, which are almost identical. However,
the actual charge distributions are by no means identical. The nucleus 
$^{48}$Ca has a larger core region of almost constant density but a 
smaller surface region, where the density drops from 90\% to 10%
\cite{Friedrich_82}. 
For the model dependent analysis it is convenient
to use Helm's model, which is known from the analysis of electron scattering
data (see \cite{Friedrich_82,Friedrich_86}) and allows to extract nuclear
extension parameters in a transparent way. In this model the nuclear density 
is parameterized \cite{Friedrich_82} by the convolution of a hard sphere 
distribution with a
Gaussian. The form factor $F_H$ is then simply given by the product of the 
form factor of the hard sphere, $F_{hs}$, with that of the Gaussian, which,
again is a Gaussian, $F_G$.
\begin{eqnarray}
\label{eq:helm}
F_H & = & F_G \cdot F_{hs}\\
F_G & = & exp(-(q\sigma )^2/2)
\end{eqnarray}
\begin{equation}
\label{eq:hard_sphere}
F_{hs} = \frac{3}{(qR_d)^2}\left(\frac{sin(qR_d)}{qR_d}-cos(qR_d)\right)\\
= \frac{3}{qR_d}j_1(qR_d)
\end{equation}
Here, $q$ is the momentum transfer and $j_1$  the first order spherical 
Bessel function. $R_d$ is the so-called 'diffraction minimum sharp radius' 
(dms-radius) and the width of the Gaussian $\sigma$ is approximately
related to the 10\%-to-90\% surface width $t_H$ via
\begin{equation}
t_H = 2.54\sigma\;\;\;.
\end{equation}
In this model, the zeroes of the form factor are determined by the zeroes of 
the Bessel function, which implies a straight forward relation between the 
dms radius and the momentum transfers $q_i$, $i$=1,2,3,... in the i-th minimum 
of the form factor.
The rms radius, which integrates over the core region and the surface zone,
is then related to the dms radius and the $\sigma$ of the distribution by
\begin{equation}
\label{eq:rmshelm}
r_{rms}=
\sqrt{\frac{3}{5}}R_d\left(1+5\left(\frac{\sigma}{R_d}\right)^2\right)^{1/2}\;\;\;.
\end{equation}
The $\sigma$ of the Gaussian can be extracted e.g. from the position $q_m$
of the first maximum (i.e. the maximum after the first minimum) of the
form factor and its magnitude $F(q_m)$ via \cite{Friedrich_82}
\begin{equation}
\label{eq:sigma}
\sigma^2=\frac{2}{q_m^2}ln\frac{3j_1(q_mR_d)}{q_mR_dF(q_m)}\;\;\;.
\end{equation}

The comparison of the dms mass radii to the dms charge radii
from electron scattering requires no correction for the proton charge radius
since multiplication of the form factor eq. (\ref{eq:helm}) with the proton 
dipole form factor has no influence on the position of the zeroes, which are 
still determined by the Bessel function. This is not the case for the width of 
the Gaussian which is expected to be larger in case of the charge form factor 
due to the contribution of the proton charge form factor.

\section{Results}
\label{sec:21}

In the following we will discuss the form factors extracted for $^{12}$C, 
$^{40}$Ca, $^{93}$Nb and $^{nat}$Pb 
\footnote{Note: In all cases targets with natural isotope composition have
been used in the experiments \cite{Krusche_02}. However, natural niobium
is mono-isotopic, and in case of carbon and calcium the admixture of other
isotopes than $^{12}$C and $^{40}$Ca is negligible at the current level
of precision.}
from the pion production data.
We will first discuss the overall features of the mass form
factors (without DWIA corrections). In the second subsection we discuss the
extraction of root-mean-square radii from the slope of the form factors
and in the final subsection the interpretation of the form factors in the 
framework of the Helm model. 

\subsection{Form factors in PWIA}
\label{ssec:30}
\begin{figure}[h]
\centerline{\resizebox{0.50\textwidth}{!}{%
  \includegraphics{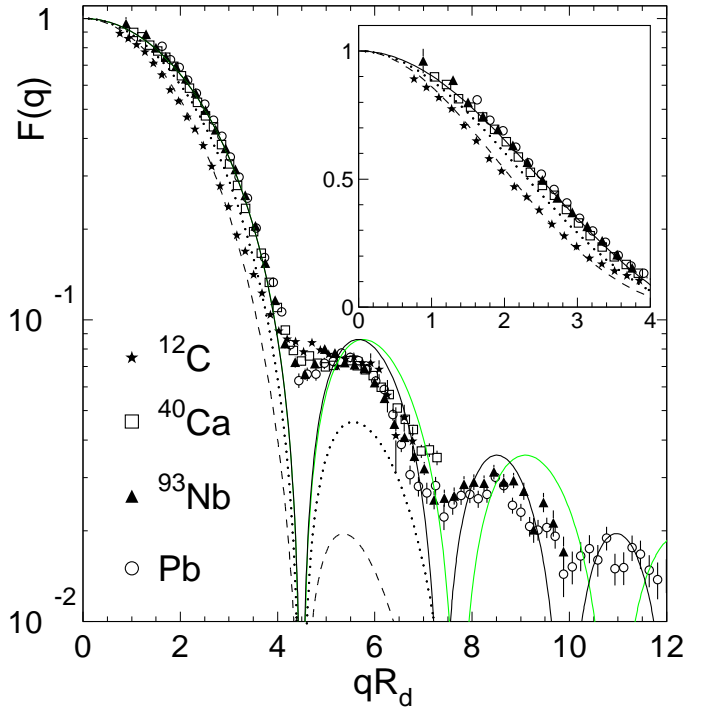}
}}
\caption{Mass form factors extracted in PWIA approximation (see text). 
Grey curve: hard sphere form factor (eq. (\ref{eq:hard_sphere})), 
solid curve: hard sphere form factor with zeroes adjusted for DWIA effects, 
dotted: hard sphere ($R_d$ of Ca) and Gaussian with $\sigma$=0.75 fm, 
dashed: hard sphere ($R_d$ of C) and Gaussian with $\sigma$=0.75 fm. 
Insert: region before first minimum in linear scale. 
}
\label{fig:pwia}       
\end{figure}

The form factors have been extracted in PWIA approximation from the cross 
section data for incident photon energies from 240 - 300 MeV with eqs. 
(\ref{eq:4}),(\ref{eq:5}),(\ref{eq:6}). At this stage 
only the overall normalization was corrected for DWIA effects. For this 
purpose the form factors were fitted at low momenta with the Taylor series 
eq. (\ref{eq:8}) and the data where divided by the fitted $a_o$ coefficient.
The results are summarized in fig. \ref{fig:pwia}, where the form factors are
plotted versus $qR_d$ which is calculated as:
\begin{equation}
qR_d(A) = q\cdot X_1/q_1(A)
\end{equation}
where $q_1(A)$ is the momentum transfer in the first diffraction minimum
and $X_1$ = 4.495 is the first zero of the Bessel function.
In this plot the positions of all diffraction minima fall on top, as one would
expect as long as the Helm model is valid, and even more the form factors for 
Ca, Nb and Pb are almost identical and agree up to the first diffraction
minimum with the form factor of a hard sphere. 
Only the carbon form factor shows the effect of a finite surface region 
in the range before the first minimum. 
The positions of the higher order minima and maxima do, however, not agree 
with the form factor of a hard sphere (grey curve), since the 
shift caused by the FSI effects is not the same for the different minima
(in case of lead roughly 3\% for the first minimum and 20\% for the 5-th
minimum). However, if the zeroes of the hard sphere form factor are
adjusted to these shifts (full curve) the data are quite well
described by the hard sphere form factor. This is surprising since 
due to the effects from the finite surface region the Helm model
predicts a significant faster fall-off of the form factors in the region
beyond the first diffraction minimum. To demonstrate this effect   
the form factors of a hard sphere convoluted with a Gaussian with 
$\sigma$=0.75 fm and the $R_d$ radii of Ca (dotted) and C (dashed) are also
shown in the figure. The predicted magnitude of the form factor
in the first maximum decreases strongly with decreasing mass number,
but the experimental results are practically identical for all nuclei.
It will be discussed in subsection \ref{ssec:32} that the effect of the 
surface thickness is canceled to a large extent by DWIA effects, so that the 
extraction of surface parameters requires a careful correction of the 
pion distortion effects.  
 
\subsection{rms-radii}
\label{ssec:31}
For the form-factor model independent extraction of the rms-radii the results
with and without correction of FSI effects (DWIA, PWIA approximation,
see fig. \ref{fig:ana}) were used. The DWIA approximation is of course 
expected to give more realistic results. However, also in this approximation 
the form factors do not exactly approach unity for $q\rightarrow 0$.

\begin{figure}[th]
\centerline{\resizebox{0.50\textwidth}{!}{%
  \includegraphics{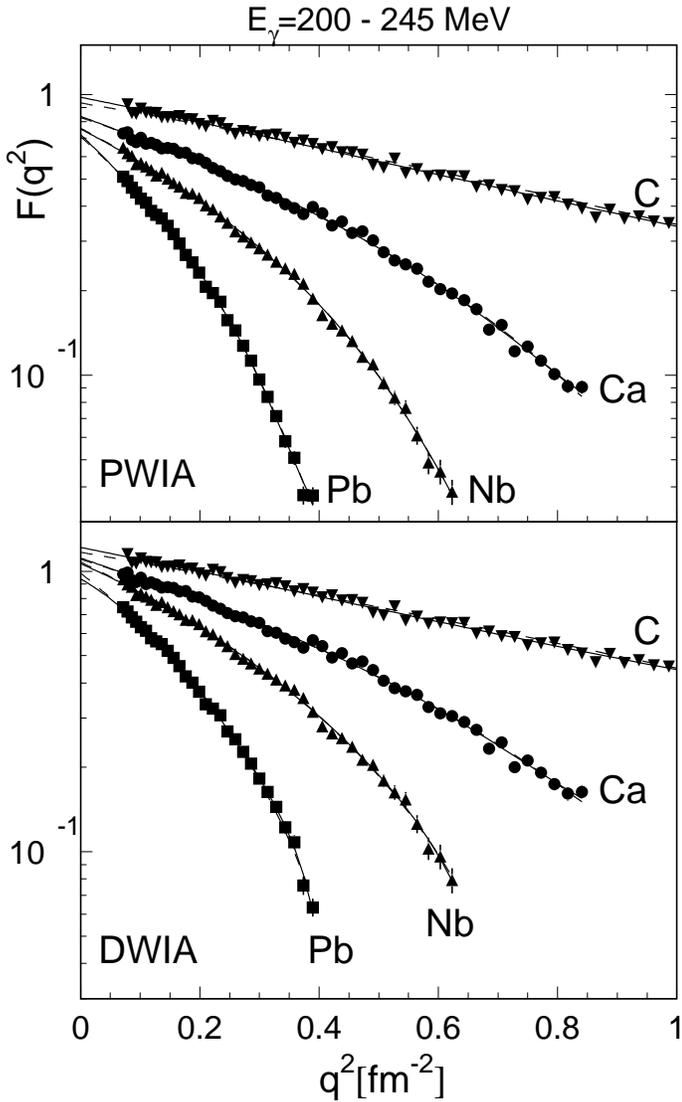}
}}
\caption{Fit of the form factors in PWIA and DWIA with polynomials of degree 
N=$4$ (solid lines) and $N$=2 (dashed lines) for incident photon energies 
between 200 MeV and 245 MeV. 
}
\label{fig:rms_1}       
\end{figure}

The size of these effects is summarized in table \ref{tab:1}, where the overall
normalization constant $a_o$ (see eq. (\ref{eq:8})) is listed for the PWIA and 
DWIA approximations for two different regions of incident photon energy. In 
all cases averages for fits with polynomials of degree $N$=2,4 are given, 
however the $N$=2,4 results differ only by a few per cent. 
The DWIA approximation brings the overall normalization closer to unity, in
particular for the higher incident photon energies. 

\begin{figure}[th]
\centerline{\resizebox{0.50\textwidth}{!}{%
  \includegraphics{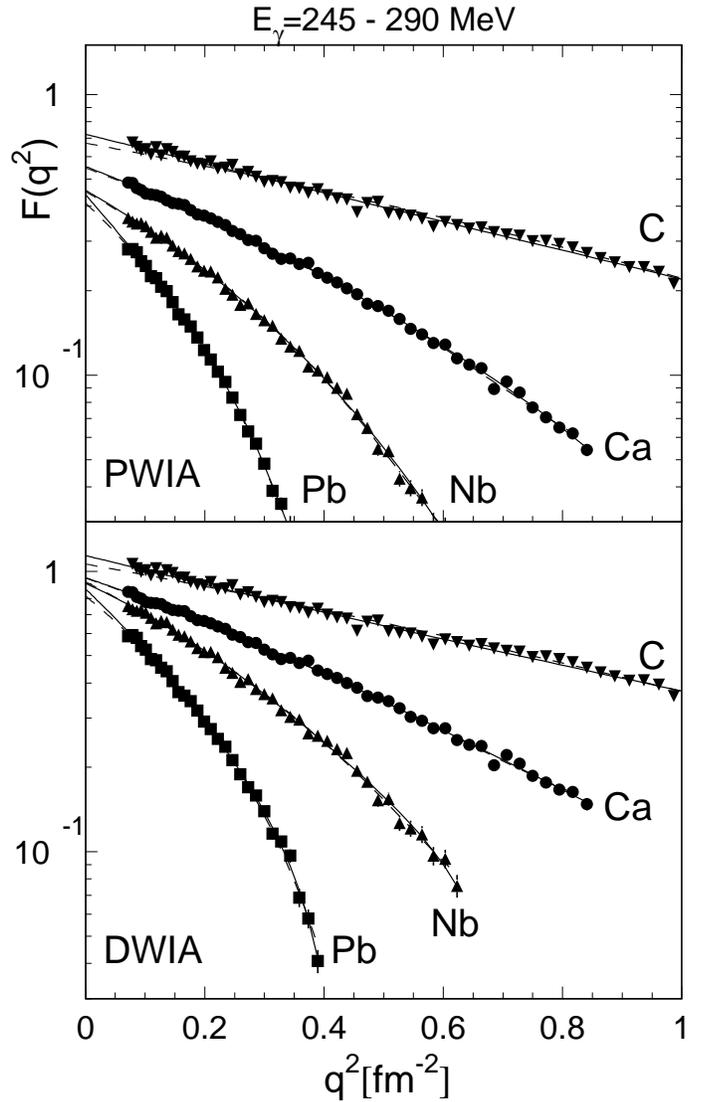}
}}
\caption{same as fig. \ref{fig:rms_1} for photon energies between 245 MeV and 
290 MeV.
}
\label{fig:rms_2}       
\end{figure}

\begin{figure}[th]
\centerline{\resizebox{0.50\textwidth}{!}{%
  \includegraphics{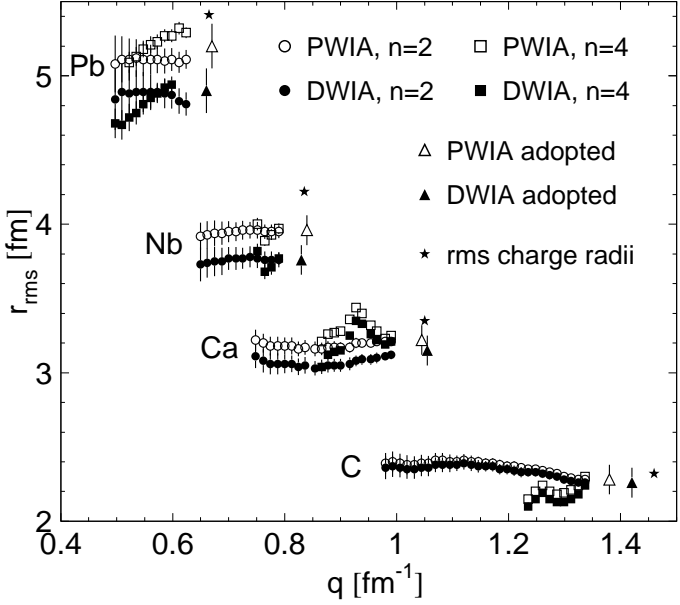}
}}
\caption{Fit results for the rms-radii extracted from 
eqs. (\ref{eq:7})-(\ref{eq:9}).
They are plotted as function of the upper limit of the fitted $q$-range for 
the PWIA and DWIA analyses and for fits with polynomials of degree $N=2,4$.
They have been averaged over the two ranges of incident photon energies.
The triangles represent the adopted values, the stars the corresponding
rms charge radii from ref. \cite{deJager_74}. 
}
\label{fig:run}       
\end{figure}

Since the rms-radii depend only on the slope of the form factor but not
on the absolute normalization (as long as $a_o$ is kept a free parameter,
see eqs. (\ref{eq:7})-(\ref{eq:9})), it is a priori not clear how large the
influence of the FSI corrections is. In order to get some estimate,
results from the PWIA and DWIA approximation are compared.
Typical fits of the data are shown in 
figs. \ref{fig:rms_1},\ref{fig:rms_2}. The results of the fits for the 
rms-radii calculated from eq. (\ref{eq:9}) are summarized in fig. \ref{fig:run}.
They are plotted as function of the upper limit of the fit range,
which allows to judge the stability of the fits. Fits have been done with 
polynomials of degree $N$=2,4. 
The fits with $N$=4 become only stable when
rather large ranges of $q$ are fitted, however in that limit they approach the
$N$=2 results. The results from PWIA and DWIA agree in most cases better
than within 10\%. However, the radii extracted in DWIA approximation are in 
all cases systematically smaller than those from the PWIA analysis, this effect 
increases with nuclear mass. 
This is expected, since the FSI effects tend to 
increase the slope of the form factor (they shift the first minimum to 
\begin{table}[bh]
\begin{center}
\caption{Normalization constants of the form factor data. Listed is the
$a_o$ coefficient of the fitted Taylor series (see eq. (\ref{eq:8})). 
The values are the average for fits with polynomials of degree $N$=2,4.} 
\label{tab:1}       
\begin{tabular}{|c|c|c|c|c|}
\hline\noalign{\smallskip}
& \multicolumn{2}{c|}{200-245 MeV} & \multicolumn{2}{c|}{245-290 MeV} \\
nucleus & PWIA & DWIA & PWIA & DWIA \\
\hline
$^{12}$C   & 0.95 & 1.20 & 0.70 & 1.10\\
$^{40}$Ca  & 0.83 & 1.11 & 0.55 & 0.95\\
$^{93}$Nb  & 0.75 & 1.07 & 0.45 & 0.92\\
$^{nat}$Pb & 0.72 & 0.99 & 0.42 & 0.83\\
\noalign{\smallskip}\hline
\end{tabular}
\end{center}
\end{table}
smaller $q$ values, see fig. \ref{fig:sabit}), which leads to an overestimate
of the radii in PWIA approximation. The adopted values of the radii are 
compared in fig. \ref{fig:run} and in table \ref{tab:2} to the nuclear charge 
radii extracted from electron scattering experiments. The uncertainties 
have been estimated from the variation of the results with the fit range 
and the agreement between the $N$=2 and $N$=4 results. 

\begin{table}[ht]
\begin{center}
\caption{Mass rms radii ($r_{rms}$) extracted in PWIA and DWIA
compared to charge rms radii ($r_{rms}^{pc}$) \cite{deJager_74}.
The charge radii have been corrected for the proton charge form factor by
subtracting in quadrature the proton rms radius ($r_p$=0.862 fm).}
\label{tab:2}       
\begin{tabular}{|c|c|c||c||c|c|}
\hline\noalign{\smallskip}
nucl. & \multicolumn{2}{c||}{$r_{rms}$[fm]} & $r_{rms}^{pc}$ &
\multicolumn{2}{c|}{$r_{rms}/r_{rms}^{pc}$} \\
& PWIA & DWIA & [fm] & PWIA & DWIA \\
\noalign{\smallskip}\hline\noalign{\smallskip}
$^{12}$C   & 2.28$\pm$0.10 & 2.26$\pm$0.10 & 2.32 & 0.98 & 0.97 \\
$^{40}$Ca  & 3.22$\pm$0.10 & 3.15$\pm$0.10 & 3.35 & 0.96 & 0.94 \\
$^{93}$Nb  & 3.96$\pm$0.10 & 3.76$\pm$0.10 & 4.22 & 0.94 & 0.89 \\
$^{nat}$Pb & 5.20$\pm$0.15 & 4.90$\pm$0.15 & 5.42 & 0.96 & 0.90 \\
\noalign{\smallskip}\hline
\end{tabular}
\end{center}
\end{table}
In all cases the mass radii are somewhat smaller than the corresponding charge
radii and this effect becomes larger for increasing mass. The effect lies
between 2 - 6 \% for the PWIA approximation and between 3 - 10 \% for the DWIA
results (note that since the DWIA correction for $^{93}$Nb was approximated
from Pb it has a larger systematic uncertainty). This result is in particular
unexpected for lead, where nuclear models (see e.g. \cite{Pomorski_97}) 
in general predict somewhat larger rms radii for the neutron distribution than 
for the protons, so that one would to the contrary expect slightly 
larger mass rms radii.
Possible explanations include still uncontrolled effects from the correction 
of the FSI in DWIA and small contaminations of the coherent cross section
with incoherent reaction processes (see sec. \ref{sec:summary}). 

\subsection{Results from the Helm model}
\label{ssec:32}
The most sensitive analysis of the form factor in the framework of the Helm
model would be a fit of the data over the full measured $q$ range. However,
it is not straight forward to apply the FSI corrections, which include
shifts of the position of the minima and modifications (reductions for most
incident photon energies) of the magnitude, consistently over the full $q$
range. Therefore, the direct determination of $R_d$ from the position of 
the minima and the determination of $\sigma$ from eq. (\ref{eq:sigma}) has 
the advantage that only specific features (shift of minima and modification 
of first maximum) of the DWIA corrections of the cross sections must be known. 
At this stage we do not attempt to fit the form factors beyond the first 
minimum, where the DWIA effects can be corrected in the way discussed in 
sec. \ref{ssec:11}. 
However, more refined treatments of the DWIA corrections are certainly
possible and will be worthwhile when more precise data become available.
Here the analysis is done in two different ways. 

\begin{figure}[h]
\centerline{\resizebox{0.50\textwidth}{!}{%
  \includegraphics{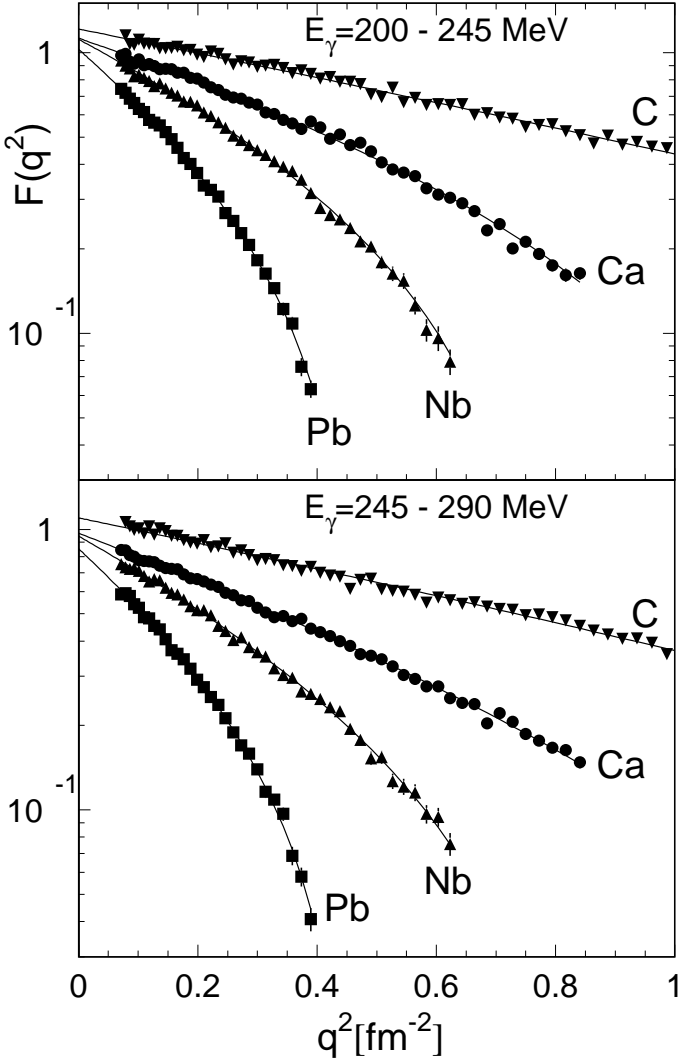}
}}
\caption{Fit of the Helm model (eq. (\ref{eq:helm})) to the form factors 
in DWIA approximation for two different ranges of incident photon energies.
}
\label{fig:helm}       
\end{figure}

\begin{table}[th]
\begin{center}
\caption{Diffraction mass radii ($R_d$) and width of the Gaussian ($\sigma$)
determined from the fit of eq. (\ref{eq:helm}) to the data. The parameters 
are averages over the two regions of incident photon energies
(200 -245, 245 - 290 MeV). The  uncertainties have been estimated from 
fit errors and the agreement between the two energy ranges.
For comparison the parameters from charge distributions $R_d^c$, 
$\sigma_d^c$ are also listed. They represent the averages of the results 
given in 
\cite{Friedrich_82} (since $^{93}$Nb is not analyzed in \cite{Friedrich_82} 
instead the average for the isotopes $^{92,94}$Zr is given in brackets). 
}
\label{tab:3}       
\begin{tabular}{|c|c|c|c|c|}
\hline\noalign{\smallskip}
nucleus & $R_{d}$[fm] & $\sigma$[fm] & $R_{d}^c$[fm] & $\sigma_c$[fm]\\
\noalign{\smallskip}\hline\noalign{\smallskip}
$^{12}$C   & 2.30$\pm$0.10 & 0.7 - 1.2 & 2.44            & 0.8   \\
$^{40}$Ca  & 3.65$\pm$0.30 & 0.6 - 1.2 & 3.79$\pm$0.04   & 0.92  \\
$^{93}$Nb  & 5.10$\pm$0.10 & 0. - 1.   & (5.04$\pm$0.04) & (0.98)\\
$^{nat}$Pb & 6.65$\pm$0.05 & 0. - 1.   & 6.66$\pm$0.04   & 0.93  \\
\noalign{\smallskip}\hline
\end{tabular}
\end{center}
\end{table}

In the first analysis the form factors in the region before the first 
diffraction minimum extracted in the DWIA approximation as discussed in 
sec. \ref{ssec:11} have been fitted with eq. (\ref{eq:helm})
multiplied with an overall normalization constant. 
The fits are shown in fig. \ref{fig:helm} for two ranges of incident photon
energies. The results for the fit parameters are summarized in tab. 
\ref{tab:3}. For the dms radii quite good agreement is found with the charge
radii extracted from electron scattering experiments. Since the fitted 
$q$-range is small for the heavy nuclei, the fits are not sensitive to the 
width of the Gaussian therefore  in case of Nb and Pb values between zero 
and unity result in similar fit qualities. 

\begin{figure}[h]
\centerline{\resizebox{0.50\textwidth}{!}{%
  \includegraphics{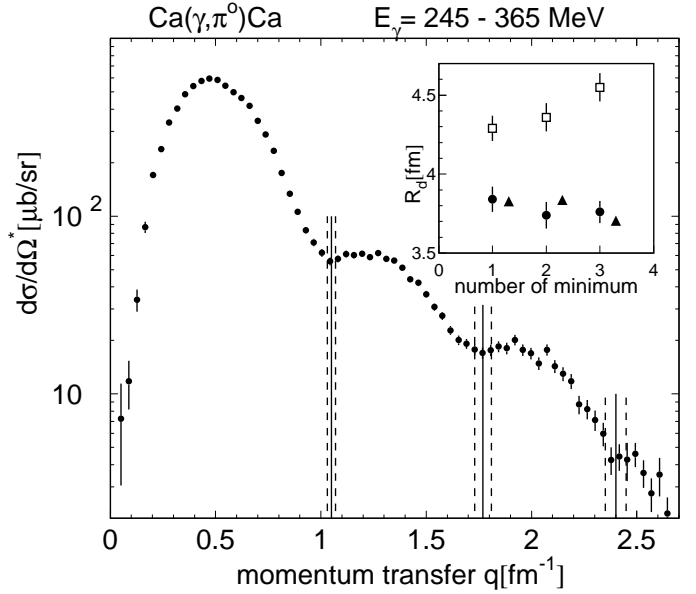}
}}
\caption{Main plot: position of the diffraction minima for $^{40}$Ca.
Insert: Extracted dms radii. Open squares: raw values, filled circles: after
correction for FSI, triangles: charge dms-radii from electron scattering
\cite{Friedrich_82}.
}
\label{fig:ca_dms}       
\end{figure}

\begin{figure}[h]
\centerline{\resizebox{0.5\textwidth}{!}{%
  \includegraphics{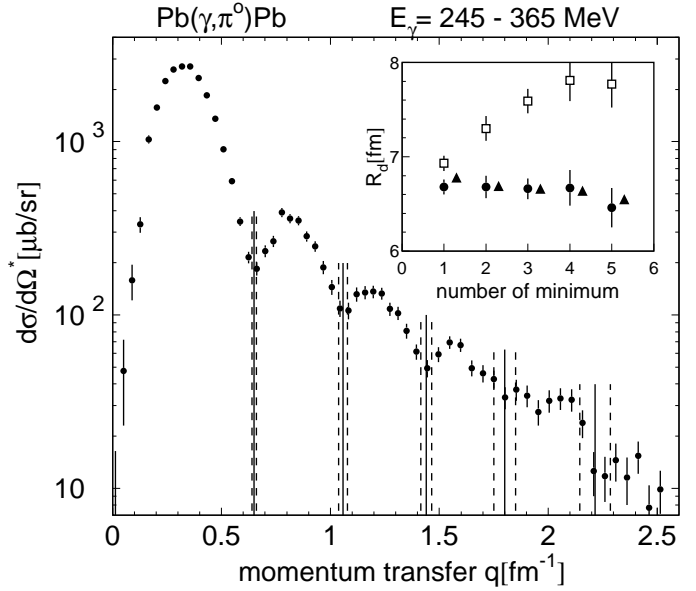}
}}
\caption{Same as fig. \ref{fig:ca_dms} for Pb.}
\label{fig:pb_dms}       
\end{figure}

A more precise determination of the radii uses the position of all observed
diffraction minima. The results for Ca and Pb are shown in figs.
\ref{fig:ca_dms},\ref{fig:pb_dms}.
For the first minimum of Ca and the first 
and second minima of Pb the positions were individually determined in bins 
of 20 MeV incident photon energy. Since no systematic trends where observed 
only the average over all bins is shown in the figures. Due to the limited
statistics the higher minima where only determined for the full energy range
from 245 - 365 MeV without considering possible energy dependent shifts. 
Although we have assigned rather conservative uncertainties to the positions
of the higher minima, their large lever arm leads to fairly precise values
of the corresponding values of the radii. 

\begin{figure}[th]
\centerline{\resizebox{0.5\textwidth}{!}{%
  \includegraphics{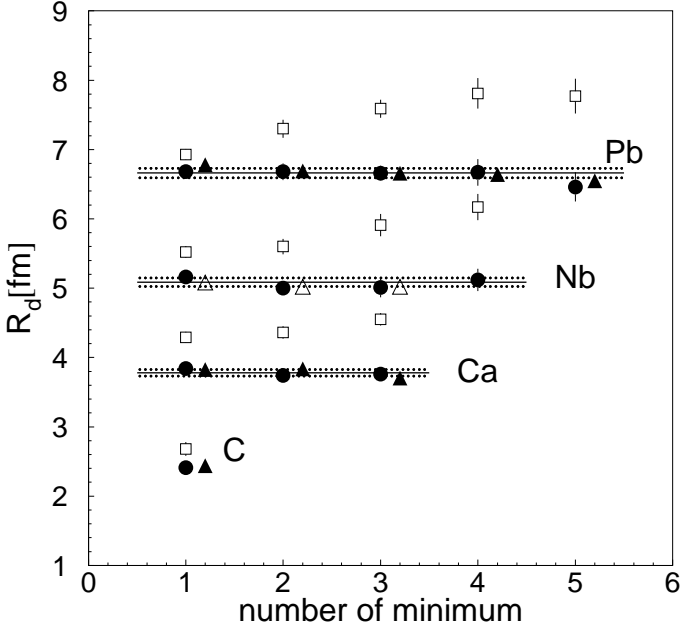}
}}
\caption{Extracted values of the dms radii. Open squares: raw values, 
filled circles: after correction for FSI, triangles: charge dms-radii from 
electron scattering \cite{Friedrich_82}. In case of Nb the open triangles 
correspond to the average of the charge dms radii for  $^{92,94}$Zr 
(values for $^{93}$Nb are not available). The full and dashed lines 
indicate average and uncertainty of the radius for each nucleus.
}
\label{fig:all_dms}       
\end{figure}

The radii have been calculated from the position of the minima via
(see eq. (\ref{eq:hard_sphere})):
\begin{equation}
\label{eq:zeroes}
R_d^{(i)}=X_i/q_i
\end{equation}
where $q_i$ is the position of the i-th minimum, $X_i$ the i-th
zero of the Bessel function and $R_d^{(i)}$ the corresponding value of the dms
radius. The results are shown in the inserts of the figures and summarized 
in fig. \ref{fig:all_dms}.

The values extracted from the raw data with eq. (\ref{eq:zeroes}) (open squares in 
the figures) show a systematic trend as function of the number of the minima.
However, this trend is completely eliminated and almost perfect agreement with
the corresponding charge radii is obtained after correction of the FSI effects
(filled circles in the figures). The shift of the position of the minima due to
FSI effects was determined from a comparison of the position of the minima in
the PWIA and DWIA calculations as discussed in sec. \ref{sec:2}
(see fig. \ref{fig:sabit}). The good agreement between the values extracted 
from the different minima after correction nicely demonstrates that the FSI
effects are well under control at least as far as the position of the 
minima is concerned.

For the charge distributions a systematic deviation of the dms radii extracted 
from the first and second diffraction minium has been found
\cite{Friedrich_82,Friedrich_86}, which was related to the depression of
the central density of the charge distribution in nuclei. It would be therefore
very interesting, to investigate if such a central density suppression 
survives also into the mass density. However, the precision achieved so far, 
does not yet allow to answer this question. The extracted 
$(R^{(1)}_d-R^{(2)}_d)$ differences are $0.10\pm 0.12$ fm, $0.16\pm 0.13$ fm, 
and $0.00\pm 0.15$ fm for $^{40}$Ca, $^{93}$Nb and Pb. This means that they are 
consistent with zero but also with the small differences observed for the 
charge radii (on the order of 0.1 fm for lead).

\begin{figure}[h]
\centerline{\resizebox{0.50\textwidth}{!}{%
  \includegraphics{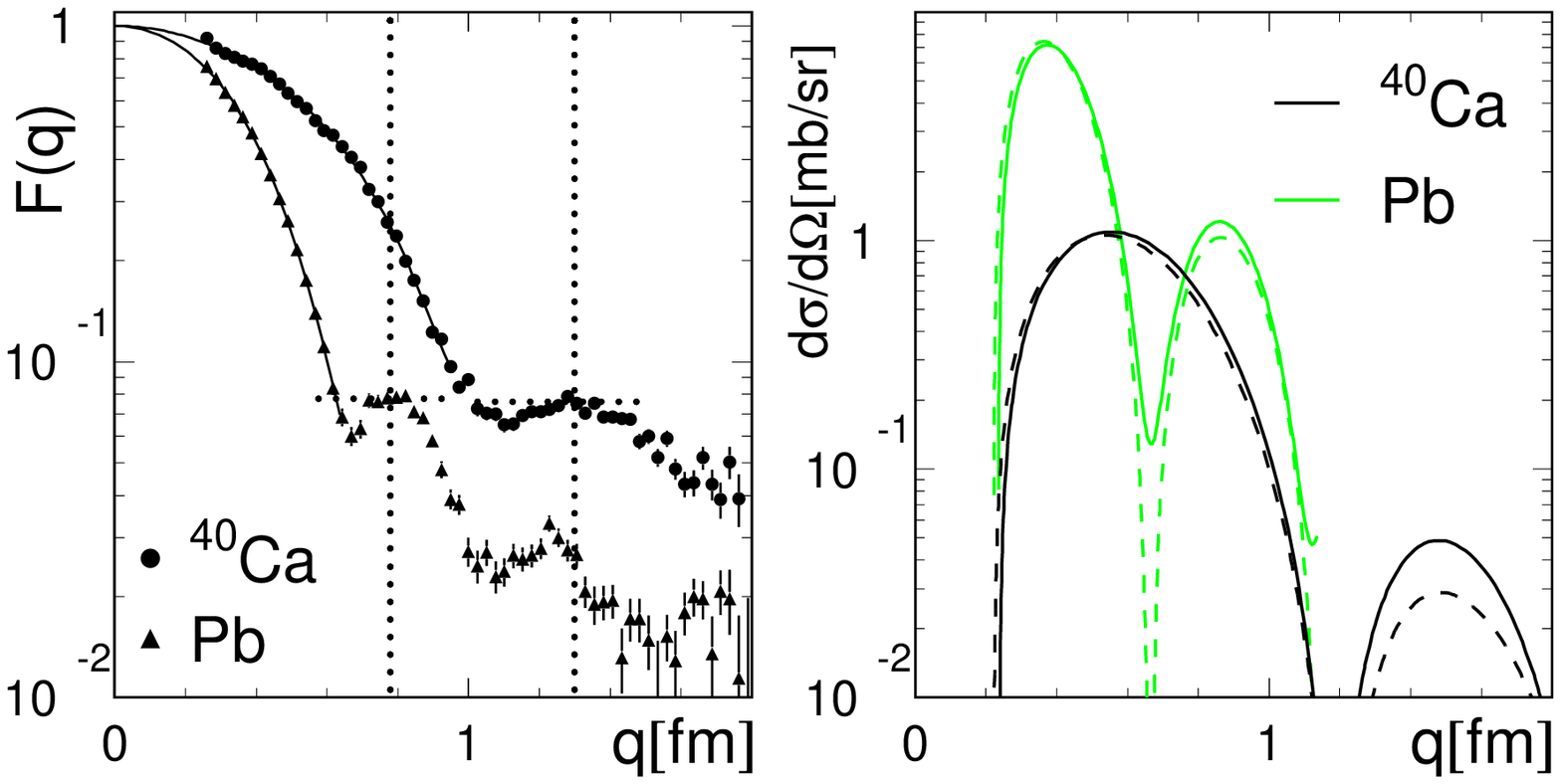}
}}
\centerline{\resizebox{0.50\textwidth}{!}{%
  \includegraphics{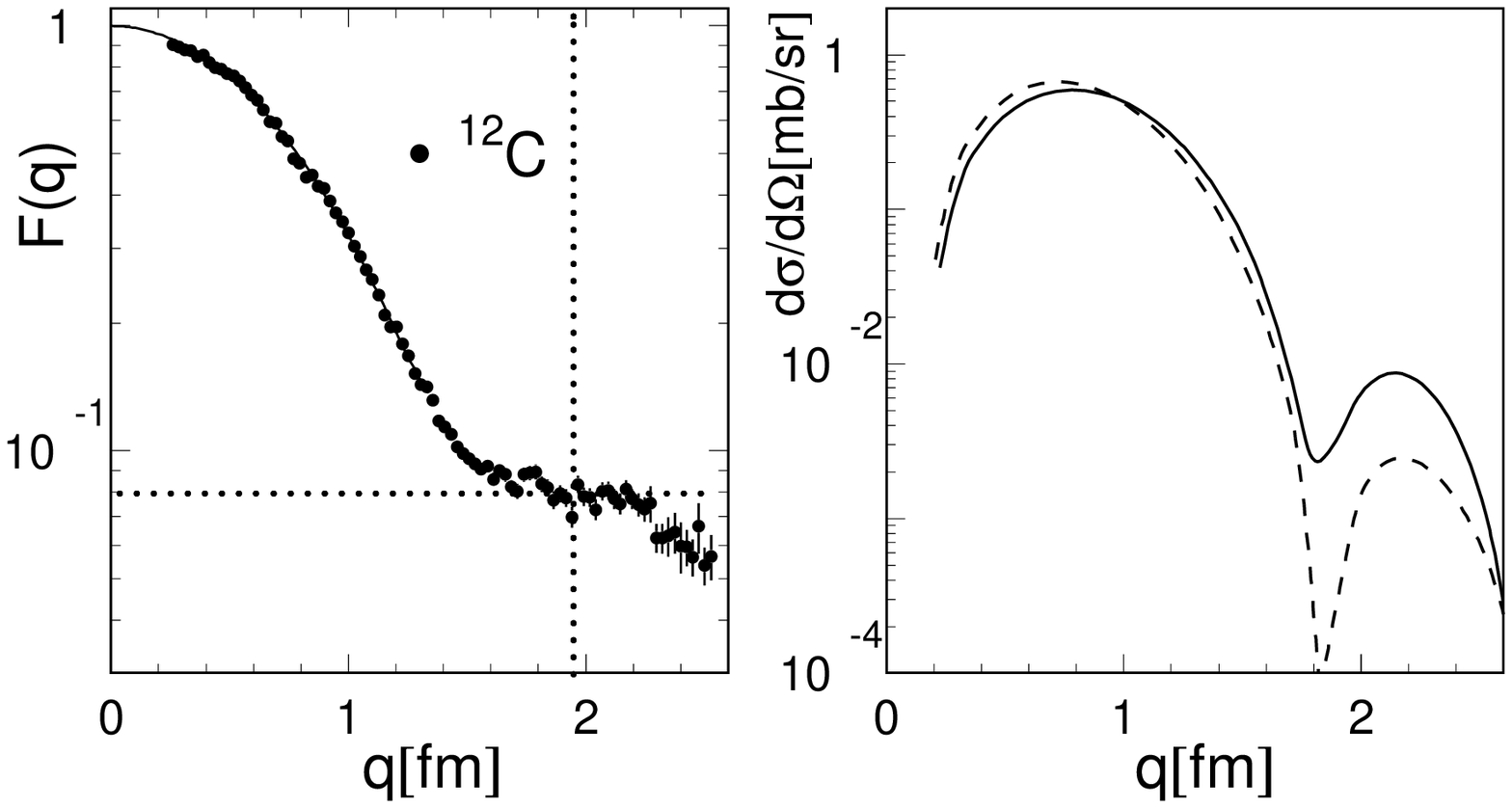}
}}  
\caption{Left hand side: position and height of the first maximum of the form
factors for carbon, calcium and lead. Right hand side: PWIA (dashed curves) and
DWIA (full curves) calculations for the differential cross sections for the
same range of incident photon energies as taken into account in the data 
analysis (200 - 260 MeV for Ca and Pb, 255 - 330 MeV for C). 
Position of the first minimum and the absolute values before the first 
minimum of the DWIA calculation are normalized to the PWIA calculation.
}
\label{fig:sigma}       
\end{figure}

Finally the first maximum of the form factors was used for a more precise
extraction of the width of the Gaussian with eq. (\ref{eq:sigma}). This
requires not only the determination of the position $q_m$ of the maximum but 
also the absolute value of the form factor $F(q_m)$, which is strongly
influenced by the FSI effects. In order to minimize the model dependency, 
the correction was done in the following way. As in sec. \ref{ssec:30} the 
form factors were fitted with Taylor series and the absolute normalization
was obtained from the condition $a_{o}$=1. The results are shown in fig.
\ref{fig:sigma}, left hand side. Position $q_m$ and magnitude $F(q_m)$
of the first maximum were determined from these data. The FSI correction
for the position was obtained in the same way as for the position of the 
minima. The correction of the magnitude was also obtained from a comparison
of the differential cross sections calculated in PWIA and DWIA.
This is shown on the right hand side of fig. \ref{fig:sigma}. 
In the picture, the position of the first minimum and the magnitude before 
the first minimum
\begin{table}[bht]
\begin{center}
\caption{Determination of $\sigma$ from the first maximum of the form factor
(see eq. (\ref{eq:sigma})).
$q_m$ and $F(q_m)$ are position and magnitude of the first maximum,
$q_m^{cor}$ and $F^{cor}(q_m)$ the same after correction for FSI effects. 
($q$ in [fm$^-1$], $\sigma$ in [fm]).}
\label{tab:4}       
\begin{tabular}{|c|c|c|c|c|c|c|}
\hline\noalign{\smallskip}
nucl. & $q_m$ & $q_m^{cor}$ & $F(q_m)$ & $F^{cor}(q_m)$  
& $q_m^{cor}R_d$ & $\sigma$\\
\noalign{\smallskip}\hline\noalign{\smallskip}
$^{12}$C   & 1.95 & 2.15 & 0.079   & 0.043  & 5.18 & 0.46 \\
$^{40}$Ca  & 1.30 & 1.41 & 0.076   & 0.058  & 5.33 & 0.53 \\
$^{nat}$Pb & 0.78 & 0.825 & 0.0776 & 0.071  & 5.49 & 0.67 \\
\noalign{\smallskip}\hline
\end{tabular}
\end{center}
\end{table}
are normalized for the DWIA calculation so that the cross sections agree in 
the $q$ range before the first minimum. The additional FSI correction for the 
magnitude of the first form factor maximum is then obtained from the square 
root of the ratio of the two calculations at this $q$ value. In this way, again 
only ratios of the model results enter into the correction factors. In the 
absence of a good DWIA calculation for Nb no analysis of the Gaussian width 
was attempted for this nucleus. The parameters of the first maximum
and the deduced values for the Gaussian width $\sigma$ are summarized in tab.
\ref{tab:4}. 

It is obvious from figure \ref{fig:sigma} and table \ref{tab:4} that the 
relative FSI corrections
(i.e. the difference of the FSI effects between the $q$ range before the first
minimum and in the first maximum) increase with decreasing mass number, 
they are almost negligible for lead but very significant for carbon 
(note that the absolute FSI effects are of course much
larger for lead than for carbon). Their effect is, that for the form factors 
normalized to unity for $q\rightarrow 0$ the magnitude of the first maximum 
is overestimated for the light nuclei in the PWIA analysis. And this effect 
cancels the influence of the finite Gaussian width on the 
magnitude in the first maximum. 

The results for the dms radii and the Gaussian width $\sigma$ are summarized 
in tab. \ref{tab:5} and compared to the corresponding values for the charge
distributions. 
\begin{table}[th]
\begin{center}
\caption{Parameters of the Helm model from the positions of the 
diffraction minima and position and magnitude of the first maximum.
The values for $R_{d}$ are the average values from fig. \ref{fig:all_dms}.
The results for $\sigma$ are from tab. \ref{tab:4}. 
$R_d^c$ and $\sigma_c$
are the respective values for the nuclear charge distributions taken 
from \cite{Friedrich_82} (the latter corrected for the finite extension
of the proton charge distribution).
}
\label{tab:5}       
\begin{tabular}{|c|c|c||c|c|}
\hline\noalign{\smallskip}
nucl. & $R_{d}$[fm] & $\sigma$[fm]  &  $R_{d}^c$[fm] 
& $\sigma_c$[fm]\\
\noalign{\smallskip}\hline\noalign{\smallskip}
$^{12}$C   & 2.41$\pm$0.09 & 0.46 &  2.44          & 0.67  \\
$^{40}$Ca  & 3.78$\pm$0.05 & 0.53 &  3.79$\pm$0.04 & 0.80 \\
$^{93}$Nb  & 5.09$\pm$0.05 & -    &  5.04$\pm$0.04 & -    \\
$^{nat}$Pb & 6.66$\pm$0.07 & 0.67 &  6.66$\pm$0.04 & 0.81 \\
\noalign{\smallskip}\hline
\end{tabular}
\end{center}
\end{table}
The mass dms-radii are in excellent agreement with their charge
counterparts, though the almost exact agreement for Ca and Pb is certainly 
by chance, given the associated uncertainties. Thus also for the diffraction 
radii there is no indication that the mass radii would be larger than the 
charge radii.

For the comparison of the width $\sigma$ with the corresponding width
$\sigma_c$ of the charge distributions one must take into
account that the values for the charge distributions include the effect 
of the finite extension of the proton, while the values extracted
from pion photoproduction refer to point like nucleons. Therefore 
$\sigma_{c}$ has been corrected for this effect with the dipole form factor
of the proton. However, even after this correction, the values extracted
for the mass distributions remain systematically smaller, in particular for the
light nuclei (see tab. \ref{tab:5}). 

\section{Summary and Conclusions}
\label{sec:summary}
Recent data for coherent photoproduction of $\pi^o$ mesons have been analyzed 
in view of nuclear mass distributions. The Helm model was used to extract 
diffraction-minimum sharp radii from the positions of the diffraction minima 
and skin thicknesses from the magnitude and position of the first maximum of 
the form factors. After corrections for FSI effects, the mass dms-radii are in 
excellent agreement with the dms-radii of the charge distributions of the 
nuclei. A stringent control of systematic effects (e.g. the FSI corrections) 
on the dms-radii is possible via the comparison of the results for different 
order minima of the same nucleus. No systematic trends have been observed. 
The uncertainties of the dms radii (3.7 \% for $^{12}$C and $\approx$ 1 \% for
all other nuclei) have reached a similar level of precision as the 
corresponding charge radii. However, the achieved precision is just not 
sufficient to exclude or establish a central depression of the mass 
distributions from the comparison of the positions of first and second
diffraction minimum.    

\begin{table}[ht]
\begin{center}
\caption{Comparison of the extracted $rms$ mass radii with charge $rms$ radii
(all values are in [fm]). $R_d$: mass dms radii (see table \ref{tab:5}),
$\sigma$: Gaussian width of the mass distributions (see table \ref{tab:5})
(value in brackets for Nb interpolated from other nuclei),
$r_{rms}^{Helm}$: rms mass radii calculated with eq. (\ref{eq:rmshelm})
from $R_d$ and $\sigma$, $r_{rms}^{PW}$, $r_{rms}^{DW}$: rms radii extracted 
from the slope of the form factors in PWIA and DWIA approximation
(see table \ref{tab:2}), $r_{rms}^c$ charge rms radii with the proton charge 
radius subtracted in quadrature.
}
\label{tab:6}       
\begin{tabular}{|c|c|c||c|c|c||c|}
\hline\noalign{\smallskip}
nucl. & $R_d$ & $\sigma$ & $r_{rms}^{Helm}$ & $r_{rms}^{PW}$ &
$r_{rms}^{DW}$ & $r_{rms}^c$ \\
\noalign{\smallskip}\hline\noalign{\smallskip}
$^{12}$C   & 2.41 & 0.46   & 2.03  & 2.28 & 2.26 & 2.32 \\
$^{40}$Ca  & 3.78 & 0.53   & 3.07  & 3.22 & 3.15 & 3.35 \\
$^{93}$Nb  & 5.09 & (0.60) & 4.08  & 3.96 & 3.76 & 4.22 \\
$^{nat}$Pb & 6.66 & 0.67   & 5.29  & 5.20 & 4.90 & 5.42 \\
\noalign{\smallskip}\hline
\end{tabular}
\end{center}
\end{table}

Root-mean-square radii of the mass distributions have been extracted in two
different ways. In the first, the slope of the form factors for 
$q\rightarrow 0$ was fitted with polynomials.  A very 
conservative estimate of the typical uncertainty due to FSI corrections of the 
shape of the form factors follows from the difference between the results 
obtained in PWIA and DWIA, which is more important the heavier the  nucleus.
In a second approach, they have been extracted in the framework of the Helm
model via eq. (\ref{eq:rmshelm}) from the dms radii and surface thicknesses.
The results are summarized in table \ref{tab:6}.
The largest discrepancies between the different analysis are in the range of 
10 \%. 

An unexpected finding is, that all results for the rms mass radii are slightly
smaller than the corresponding rms charge radii, even after the latter have 
been corrected for the proton charge radius. Since at least for nuclei like
$^{208}$Pb models predict slightly larger rms radii for the neutron 
distributions than for the proton distributions one would have expected the 
opposite. At the same time the skin thicknesses extracted from the Helm model
are also systematically smaller than their charge counter parts. In a sense,
these two effects are consistent since smaller skin thicknesses combined
with identical diffraction radii will lead to smaller rms-radii.
However, as yet it is not clear if these effects are real or if they can be 
explained by so far not understood systematic effects in the data or in the 
model calculations used for the DWIA corrections. A possible explanation
could be a small remaining incoherent background component in the cross section
data. The angular distributions of incoherent $\pi^o$ photoproduction 
involving excited nuclear states peak at larger angles than the coherent 
reaction. Therefore, such background, which so far is only 
suppressed by kinematical cuts, would tend to decrease the slope of
the form factor at small $q$, it would enhance the magnitude of the first
diffraction maximum, but it would not change the position of the diffraction 
minima. In such a scenario, rms-radii and skin thicknesses would be 
underestimated, but the dms-radii would not be effected. 
Clearly, further improvements on the experimental side and for the model
calculations are necessary and possible. On the experimental side, large
improvements in the statistical quality of the data are possible with the
now available 4$\pi$ electromagnetic calorimeters. Furthermore, improvements
in the energy and angular resolution for the detection of the $\pi^o$
decay photons will allow an even more stringent suppression of incoherent
pion production reactions than achieved in \cite{Krusche_02}. The
background situation will furthermore very significantly improve since with 
the large solid angle coverage the detection of decay photons from excited 
nuclear states will not only allow to veto such events much more efficiently 
than previously, it will actually allow a detailed investigation of the 
incoherent processes  (which are very interesting in their own right) so that 
any remaining background can be subtracted. Furthermore, a more systematic
investigation of coherent pion photoproduction from many nuclei in the 
framework of models is clearly desirable.

\section{Acknowledgments}
First of all I like to thank J. Friedrich for many stimulating 
discussions, valuable suggestions, and detailed comments to the manuscript. 
I gratefully thank S.S. Kamalov for the
provision of the DWIA calculations. This work was supported by 
Schweizerischer Nationalfonds.

\end{document}